\documentclass[runningheads,a4paper]{llncs}
\usepackage{graphicx}
\usepackage{wrapfig}
\usepackage{multirow}
\usepackage{todonotes}

\usepackage[binary-units]{siunitx}
\usepackage{subcaption}
\usepackage{tikz}
\usetikzlibrary{
    calc,
    shapes,
    spy,
}

\usepackage[hidelinks]{hyperref}
\usepackage[capitalise]{cleveref}

\begin{document}
\title{
On the Non-Associativity of Analog Computations
}

\newcommand{\repeatthanks}{\textsuperscript{\thefootnote}}


\author{Lisa Kuhn\thanks{These authors contributed equally.}\href{https://orcid.org/0000-0001-9901-4995}{\includegraphics[scale=0.08]{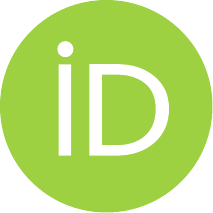}}, \and
Bernhard Klein\repeatthanks\href{https://orcid.org/0000-0003-0497-5748}{\includegraphics[scale=0.08]{img/orcid.pdf}},
Holger Fröning\hspace{0.5mm}\href{https://orcid.org/0000-0001-9562-0680}{\includegraphics[scale=0.08]{img/orcid.pdf}}}

\authorrunning{L. Kuhn et al.}

\institute{Institute of Computer Engineering, Heidelberg University, Germany
\email{lisa.kuhn@stud.uni-heidelberg.de}\\
\email{\{bernhard.klein,holger.froening\}@ziti.uni-heidelberg.de}}

\maketitle            

\begin{abstract}

The energy efficiency of analog forms of computing makes it one of the most promising candidates to deploy resource-hungry machine learning tasks on resource-constrained system such as mobile or embedded devices.
However, it is well known that for analog computations the safety net of discretization is missing, thus all analog computations are exposed to a variety of imperfections of corresponding implementations.
Examples include non-linearities, saturation effect and various forms of noise.
In this work, we observe that the ordering of input operands of an analog operation also has an impact on the output result, which essentially makes analog computations non-associative, even though the underlying operation might be mathematically associative.
We conduct a simple test by creating a model of a real analog processor which captures such ordering effects.
With this model we assess the importance of ordering by comparing the test accuracy of a neural network for keyword spotting, which is trained based either on an ordered model, on a non-ordered variant, and on real hardware.
The results prove the existence of ordering effects as well as their high impact, as neglecting ordering results in substantial accuracy drops.

\end{abstract}

\section{Introduction}

Machine learning (ML) is one of the key technologies of this century, and the main contributing factor for recent advances in computer vision, natural language processing, speech recognition and signal processing.
However, it comes at tremendous costs in terms of time and energy, which is conflictive with a pervasive deployment of ML-based processing on resource-constrained computing systems, such as mobile, handheld, wearable and similar devices.
As the diminishing returns from improvements in CMOS technology result in challenges in sustained performance scaling, new avenues for a successful performance scaling have to be found.
One promising option are analog forms of computation, where instead of a discretization of electrical quantities such as current or voltage, computations are directly performed on such quantities.
In theory, this is governed by Kirchhoff's circuit laws.
If we now consider Kirchhoff's current law, which states that the sum of currents flowing into a node is equal to the sum of currents flowing out of this node, we note that  according to this law all current is preserved, thus the implementation of an add operation based on the sum of currents does not require any energy.
Similarly, a multiplication can be represented by a current pulse, in which the two input operands are represented by the magnitude and the length (time) of this pulse, respectively.
Again, in theory, implementing a multiplication based on such a representation does not require any energy.
In practice inductance, resistance and thermal noise dissipate small amounts of power.
Nonetheless for workloads which can tolerate lower precisions, analog computations are considered superior to their digital complement in terms of energy efficiency~\cite{9197673}.

While digital CMOS is based on discretization to model the problem being solved, analog (electrical) computing~\cite{8278135} leverages continuous electrical quantities, such as voltage or charge.
While this greatly improves energy efficiency and speed, it comes at the cost of long-term and short-term variations, such as noise, crosstalk, and manufacturing process variations.
As a result, finding remedies and suitable design points that trade among efficiency and robustness is mandatory~\cite{9197673,4633828}.
Related work on analog computing is due to its unsafeness rather application-specific. 
Murmann proposes a mixed-signal processor architecture for neural network architectures~\cite{9197673}, which proposes an extremely dense and energy-efficient processing array.
Another analog computing platform for ML applications is BrainScaleS-2 (BSS-2) as described by Schemmel et al.~\cite{schemmel2021accelerated,4633828}, which utilizes analog operations for the computation of spiking as well as artificial neural networks.
Also, resistive-switching random access memory (RRAM)~\cite{Wong2012} is often used in an analog fashion to implement the dot product of neural networks in an array of resitive elements, where resistance representing the weight of a connection of a neural network. 
Similar to analog computations in CMOS, resistive memory is also vulnerable to various imperfections, including for instance thermally activated drift and diffusion.
Notably, all previous examples focus on neural networks and are crossbar-based matrix-vector architectures.
There is also work on process-oriented applications such as ordinal or partial differential equations~\cite{cowan2006}, providing further evidence about the energy efficiency of analog computations.

While various sources of imperfections of such analog computations have already been explored, in particular including noise, CMOS-based analog computing can additionally be vulnerable to ordering effects.
Consider the dot product of two input vectors $\textbf{a}$ and $\textbf{b}$ as the key operation of neural networks: $c=\sum_i(a_i\cdot b_i)$.
Usually the associativity of the add operation guarantees that this iterator can be traversed in arbitrary order\footnote{Mathematically true, but in computing this guarantee is usually relaxed due to data types such as floating point, which, however, is only noticeable in rare cases.}.
With regard to CMOS-based analog computing, however, we hypothesize that the ordering of input operands has effects on correctness of the result of this dot product.
\begin{wrapfigure}[22]{R}{0.45\textwidth}
		 \centering
		 \vspace{-2\baselineskip}
		 \includegraphics[width=1.0\linewidth]{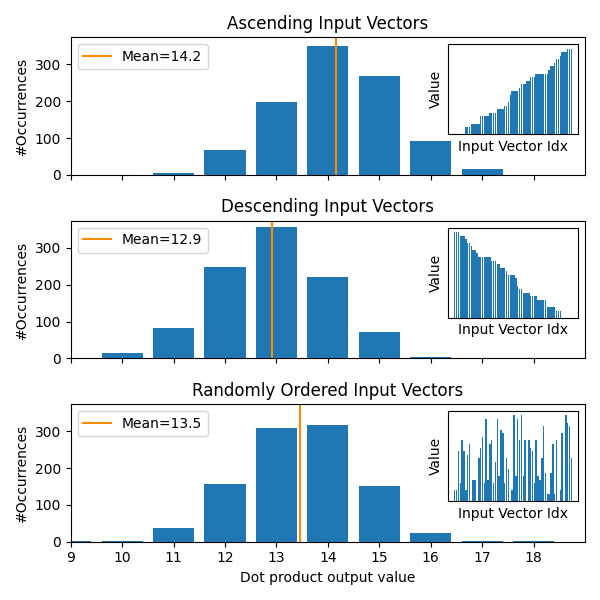}
		 \caption{Distribution of output values for an analog computation of the dot product $c=\textbf{a}\cdot \textbf{b}=\sum_i a_i b_i$ with differently-ordered input operands \textbf{a} and \textbf{b}. The inset images illustrate the ordering of values within the input vector.}
		 \label{fig:input_ordering}
\end{wrapfigure}

For instance, \emph{saturation of inputs} can occur if those inputs are of high magnitude and arrive in short time intervals, as the intermediate time in between multiple inputs is then not sufficient to return to an unbiased state.
Similarly \emph{membrane leakage} can have similar effects, as here different charges are accumulated but one cannot assume its leakage current to be zero, thus charges will naturally fade over time.
Both effects highly depend on the sequence of input values.
We report the result of a simple experiment with different orderings of input values in Figure~\ref{fig:input_ordering}, which was conducted on the BSS-2 analog processor~\cite{schemmel2021accelerated}.
As on can see, even for such a limited setting there is a huge variance in the resulting output.

We thus propose to include ordering effects when modeling such analog processors.
Such an accurate model can be of various help, as it allows computations without access to the hardware device, for instance when training or fine-tuning a neural network.

In detail, we propose an assessment of importance of such ordering effects by modeling analog hardware in the form of a model architecture based on a \emph{set of transformers}, thereby relying on the expressive power of transformers to represent sequences.
Besides the model architecture, we create a training and validation data set by a variety of measurements to capture effects such as non-linearities and saturation effects as well as the inherent ordering effects present in an analog processor.
Then, this combination of data and model should be able to represent all static imperfections.
To represent the stochastic electrical noise, we add artificial noise sampled from a Gaussian distribution, calibrated with noise patterns from real hardware measurements.
In a second test case, we permute all input operands to create an ablation test case, the \emph{non-ordered transformer set}, in which the ordering information is removed.
The trained hardware models are then used for keyword spotting based on the Google Speech Commands (GSC) data set~\cite{warden2018speechcommands} as an exemplary workload. 
In particular, we compare to previous results from \cite{klein2021item}, in which a similar objective was pursued but without considering ordering effects, and to training on real analog hardware.

In summary, this work makes the following contributions:
\begin{enumerate}
	\item 	Proposing a \emph{transformer-set}-based assessment method for the importance of ordering.
	\item 	Creating suitable data sets for training and validation, with and without ordering information.
	\item 	Investigating the importance of ordering on an exemplary learning task.
\end{enumerate}

This work extends previous efforts \cite{klein2021item}, which we refer to in the following as \emph{white-box model} as it is based on parametric components such as lookup tables, splines and additive noise.
This model mainly considered noise and static variations as source of imperfections for analog computations, in particular in comparison to digital ones. 
However, capturing those imperfections in the model and using it for the training of a neural network architecture was not able to meet the quality of training such a network directly on the analog hardware, which is usually referred to as \emph{hardware-in-the-loop} training.
Thus, while the previous work substantially narrowed the gap between plain training and \emph{hardware-in-the-loop} training, a notable gap was remaining.
One of our key questions is thus, by how much will this gap be closed by also incorporating ordering effects due to the non-associativity of analog computations, or, in other terms, how much closer will the updated model be to reality.

The remainder of this work is structured as follows: 
we continue with providing necessary background as well as a brief review of related work in the next section.
Section~\ref{sec:methodology} presents the overall methodology in more detail, in particular with regard to ordering effects and how a \emph{transformer set} can be used for an assessment of the corresponding importance.
Section~\ref{sec:results} reports training methods and test results, 
while the last sections summarizes key finding and provides an outlook.

\section{Background and Related Work}

\subsection{BrainScaleS-2}\label{subsec:bss2}

\begin{wrapfigure}[16]{R}{0.45\textwidth}
        \vspace{-2\baselineskip}
        \centering
        \resizebox{0.5\textwidth}{!}{\definecolor{blue}{HTML}{1f77b4}%
\definecolor{red}{HTML}{d62728}%
\definecolor{green}{HTML}{2ca02c}%
\definecolor{orange}{HTML}{ff710e}%
\definecolor{yellow}{HTML}{fee23e}%
\definecolor{cadc}{HTML}{1f77b4}%
\definecolor{input}{HTML}{ff7f0e}%
\definecolor{hidden}{HTML}{2ca02c}%
\definecolor{output}{HTML}{555555}%
\tikzset{block/.style={font={\rmfamily\footnotesize},align=center}}%
\tikzset{box/.style={draw=black!90}}%
\tikzset{block label/.style={fill=white,font={\rmfamily\footnotesize},inner sep=0.05cm}}%
\tikzset{%
	neuron/.style = {%
		draw=black,%
		circle,%
		inner sep=0pt,%
		minimum width=0.4cm%
	},%
	driver/.style = {%
		minimum height=0.45cm,%
		draw=black,%
		regular polygon,%
		regular polygon sides=3,%
		shape border rotate=-90,%
		inner sep=0pt%
	},%
}%
\begin{tikzpicture}[
	scale=0.8,
	>=stealth,
	transform shape,
	line width=1.0\pgflinewidth,
	anchor=center,
	spy using outlines=circle,
]
	\pgfdeclarelayer{background layer}
	\pgfsetlayers{background layer,main}
	\draw[use as bounding box,inner sep=0pt,draw=none] (0.0,0.0) rectangle ++(7.0,4.5);

	\begin{scope}[scale=1.15]
		\foreach \x in {0,1,...,6} {

			\node[neuron,output,thick] (nrn \x) at (0.8 + \x*0.5,0.35) {};
			\draw (nrn \x.north) ++ (0.0,0.01) -- ++(0.0,2.5);

		}

		\foreach \y in {0,1,...,4} {
			\node[driver,thick] (drv \y) at ($(nrn 0) + (-0.5,0.5 + \y*0.5)$) {};
			\draw (drv \y.+45) -- ($(drv \y.center) + (0.15, 0.125)$) -- ++(3.5,0.0);
			\draw (drv \y.-45) -- ($(drv \y.center) + (0.15,-0.125)$) -- ++(3.5,0.0);

			\foreach \x in {0,1,...,6} {

				\fill[hidden] (drv \y -| nrn \x) ++ (0.0, 0.125) circle (0.05cm);
				\draw[white] (drv \y -| nrn \x) ++ (0.0, 0.125) ++ (-0.045,0.0) -- ++(0.09,0.0);
				\draw[white] (drv \y -| nrn \x) ++ (0.0, 0.125) ++ (0.0,-0.045) -- ++(0.0,0.09);
				\fill[hidden] (drv \y -| nrn \x) ++ (0.0,-0.125) circle (0.05cm);
				\draw[white] (drv \y -| nrn \x) ++ (0.0,-0.125) ++ (-0.045,0.0) -- ++(0.09,0.0);
			}
		}

		\node[rectangle,thick,draw,cadc,inner sep=2pt,minimum width=3.5cm] (cadc) at ($(nrn 2) + (0.5,3.0)$) {\fontsize{6}{6}\selectfont Columnar ADC};
		\foreach \x in {0,1,...,6} {
			\draw[blue] (nrn \x.55) -- ++(55:0.12) coordinate (tmp) -- (tmp |- cadc.south);
		}

		\node[rectangle,thick,draw,inner sep=2pt,minimum width=3.5cm,above=0.07cm of cadc] (ppu) {\fontsize{6}{6}\selectfont SIMD Processor};

	\end{scope}

	\coordinate (sherlock) at (5.6,2.2);
	\spy[draw,height=1.2cm,width=1.2cm,magnification=2.5,connect spies] on (nrn 6 |- drv 0) in node at (sherlock);
	\node[align=center] at ($(sherlock) - (0.0,1.1)$) {\fontsize{7}{7}\selectfont signed\\[-0.3em] \fontsize{7}{7}\selectfont synapse};
\end{tikzpicture}}
        \caption{Schematic of the BSS-2 analog core, with synapse drivers (triangles), neurons (large circles), and synapses (small green circles)~\cite{cramer2020training}.}
        \label{fig:bss2_schematic}
\end{wrapfigure}
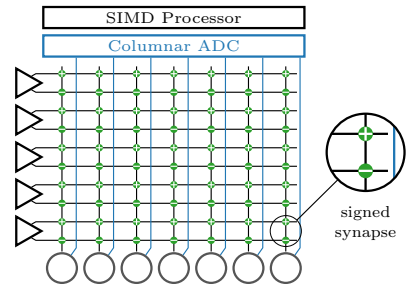

BrainScalesS-2 is a mixed-signal neuromorphic ASIC, manufactured in \SI{65}{\nano\meter} CMOS\@ technology~\cite{schemmel2021accelerated} and supports the execution of spiking as well as artificial neural networks within its analog core~\cite{weis2020inference}.
Moreover, it is an ideal candidate to characterize typical challenges that have to be overcome to transfer and execute digital algorithms successfully on an analog matrix-multiply accelerator.
Figure~\ref{fig:bss2_schematic} illustrates the crossbar structure of BSS-2: 512 neuron circuits each connected to a column of 256 synapses and to a ADC to readout the results.
To allow signed weights, two hardware synapses are combined to a single virtual synapse.

\begin{figure}[t]
	\centering
	\includegraphics[width=1.0\linewidth]{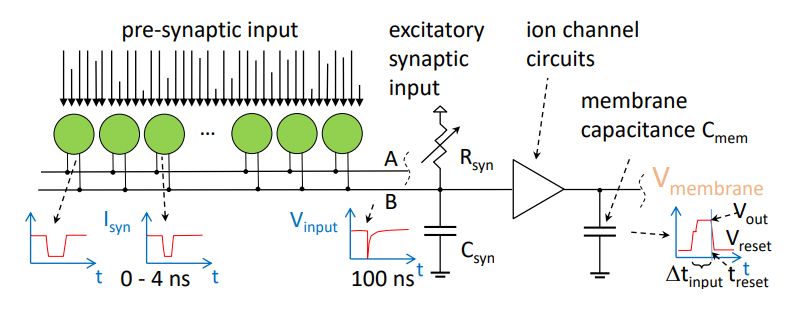}
	\caption{Detailed view of a column, illustrating the integration of synaptic pulses to one neuron with BSS-2 in non-spiking mode~\cite{schemmel2021accelerated}.}
	\label{fig:bss2_detailed}
\end{figure}

Fundamentally BSS-2 sends activation vectors to the \emph{synaptic drivers}, and converts them to corresponding time frames.
The weights are represented as currents and thus, a synapse multiplies an activation with a weight value with a corresponding current pulse.
Those charges $Q=I \cdot t$, amplified by an operational transconductance amplifier (OTA), are accumulated on the neuron's membrane capacitance and finally converted back to digital with an 8-bit ADC.
After the computational results are read the membrane capacitors are reset to default potential and the next operation can be executed. 
Figure~\ref{fig:bss2_detailed} illustrates this process in detail.

There are multiple sources of uncertainty involved.
The circuitry is not perfectly linear throughout its full dynamic range; various saturation effects are observable which further lead to non-linearities.
Additionally, due to imprecisions of the manufacturing process the ASICs differ and each chip is individual.
Calibration methods are used to achieve an mostly uniform and linear behavior of the chip.
Nonetheless, the individual chip instances differ in their components, calibration and overall behavior.
Moreover, fundamentally any analog computation is fraught with electrical noise of various sources, including thermal noise and inducted noise from neighboring circuitry.

To exploit the energy-efficiency superiority of analog computing, better methods to tolerate their peculiarities are required.
Understanding the specific behavior of analog accelerators is therefore a central step to develop the technology to majority and make it suitable for real world problems.

\subsection{Transformers}
Ever since the introduction by Vaswani et al. \cite{Vaswani2017}, transformers have become a popular and successful model in a variety of deep learning applications. 
Originally consisting of an encoder-decoder structure for sequence-to-sequence tasks like machine translation, the encoder is also applied successfully on its own in classification/regression settings where only a representation of the input is necessary, whereas the decoder is useful for generative applications. 

The main building block of transformers consists of a (multi-head) self-attention module and a following feed-forward network. To build deeper models, this block is stacked along with residual connections and layer normalization after each block. 
Since the main Transformer components (self-attention \& feed-forward networks) are permutation equivariant \cite{Lin2021}, it is necessary to explicitly incorporate information about the sequence order into the model, e.g. via sinusoidal position encodings or learned position embeddings \cite{Devlin2018} of the input.
Overall, transformers are able to relate different positions of the input sequence and to model long-range dependencies via self-attention and are therefore a flexible and powerful tool to learn from and represent ordered sequences. 

\subsection{Related work}

For analog forms of computations, there exists a plethora of related work, with recent selected works including proposals by Murmann and Schemmel, respectively~\cite{9197673,schemmel2021accelerated}.
The research body on the (non-)associativity of computations has received much less attention.
Best to our knowledge, there is yet no treatment of associativity effects for analog computations.

However, principally floating-point operations are also not associative, at least not beyond the scope of a single instructions.
In this regard, for a review of properties and pitfalls of floating-point arithmetic, we refer the reader to the excellent work by Goldberg~\cite{Goldberg1991}.

Other previous works have studied the effects of numerical precision in scientific computing, and found in particular large summations and accumulation operations are susceptible to such effects~\cite{BAILEY201210106}.
In more detail, it is reported that repeated operations as found in loops as error-prone due to accumulation effects.
As loops are often the foundation of parallelization, naturally parallel forms of execution are thus also susceptible to error accumulation.
The resulting consequences regarding verifying computations based on floating-point arithmetic are discussed in \cite{10.1145/1353445.1353446}.
Similarly, \cite{osti_976992} reports effects of non-associativity for large-scale numerical computations, and notably highlights the need for reproducibility in spite of dynamically scheduled systems such as multi-threaded systems.
An algorithm on parallel but reproducible summation based on floating-point has been proposed by Demmel in \cite{6875899}.
Some work proposes to use ML techniques based on neural networks to correct such sub-precision errors, in \cite{HARIDAS2022100081} on the example of computational fluid dynamics.

Essentially, we conclude the review of related work that there are similarities with the non-associativity of floating-points, but sources are majorly different (numerical precision vs. dynamic effects), and also the scale of the resulting effects seems to be quite different.
With regard to the latter, one can observe that a substantial amount of repetition is required to unveil non-associativity effects of floating-point, while our initial experiments for similar effects of analog computation only require a couple of dozens repetitions.

\section{Methodology}
\label{sec:methodology}

\subsection{Ordering Problems of Analog Computing}

The causes of order dependencies in analog calculations are as diverse as the possibilities for implementing them.
Nevertheless, BSS-2~\cite{schemmel2021accelerated} is a good, representative example of a crossbar-matrix-vector accelerator which integrates charges in the form of current pulses.

\paragraph{Saturation of synaptic inputs} can have a major impact to neighbors in the activation vector and by this break associativity.
Normally the resulting charge of a synapse decreases the voltage of the line.
Then the OTA compares the potential of the line with a reference potential and adds a proportional current to the membrane capacitor, as illustrated in Figure~\ref{fig:bss2_detailed}.
However, if several strong signals arrive in a short time interval, the potential of the line does not recover before the next synaptic charge is injected.
This way, the electrical potential can decrease to the non-linear regime and thus the wrong amount of charge is integrated on the membrane capacitor.
Consequently a burst of large values can negatively impact the accuracy of the calculation. 
A more detailed discussion of those effects can be found in~\cite{weis2020ma}.

\paragraph{Synaptic driver variations} occur due to the manufacturing process of the hardware, together with changes due to calibrations, which itself focus mainly on overall linear behavior of the chip. 
These variances are driver-specific and therefore dependent on the position in the activation vector.
Even so, this kind of static synapse-specific effects are covered by the lookup-table of the \emph{white-box model}~\cite{klein2021item}, the position-specific variances can fundamentally impact the associativity of operations. 

\paragraph{Membrane leakage} is another time and therefore order-dependent effect.
If the membrane leakage current is not negligible, the integrated charge fade over time.
And with this, the first results of the calculation are more effected by the leakage.
The balance of the dot product components---first to last---impacts the final sum and can break associativity.

\subsection{Modeling the Hardware with a Set of Transformers}

\begin{wrapfigure}[20]{r}{0.3\linewidth}
		 \centering
		 \includegraphics[width=0.3\textwidth]{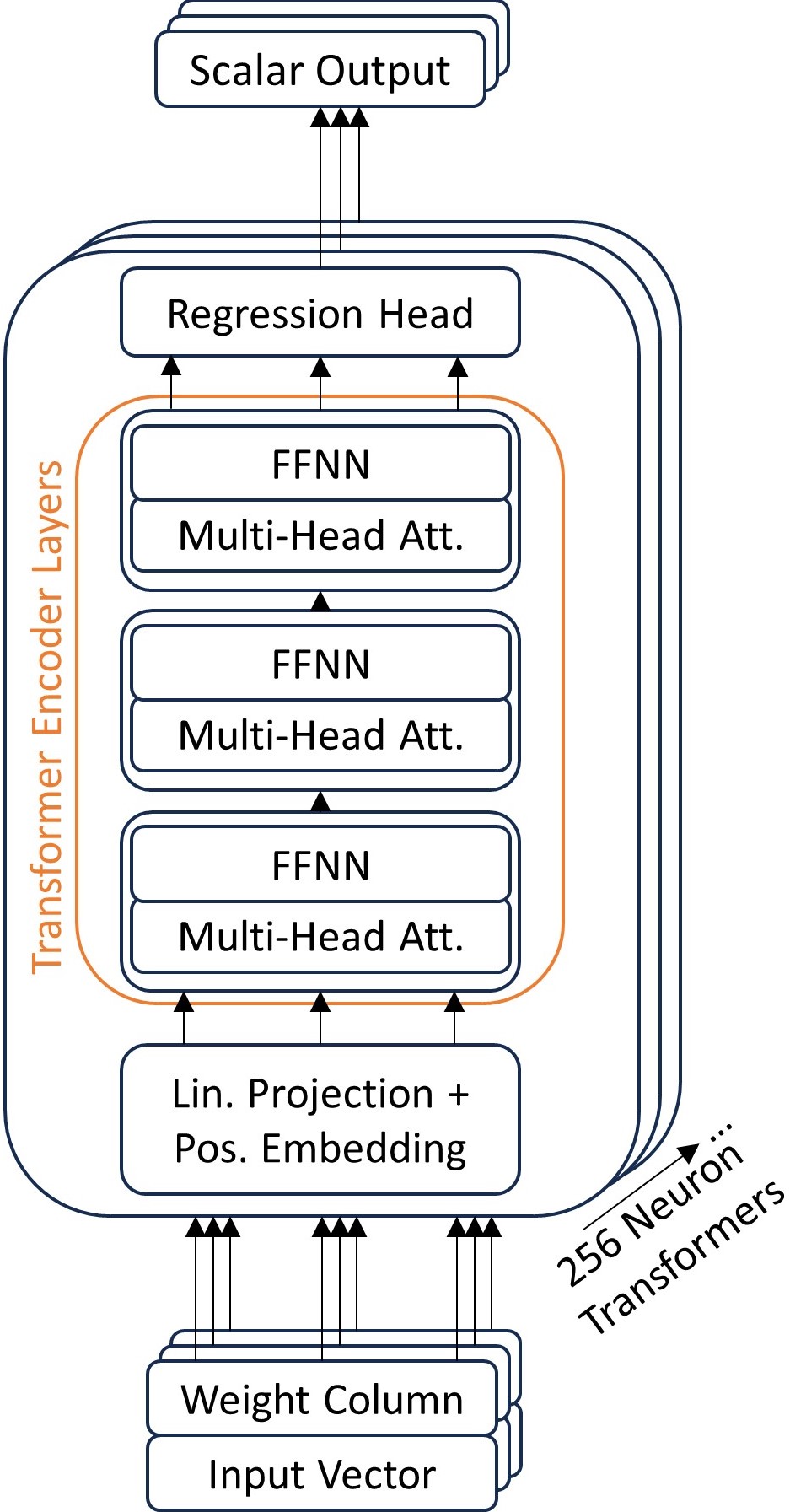}
		 \caption{Schematic view of the \emph{Transformer set}}
		 \label{fig:transformer_schema}
\end{wrapfigure}

With the design of the \emph{white-box model}~\cite{klein2021item}, we observed that modeling the characteristics of an analog matrix-multiply accelerator can be challenging.
In this work, a very generic model with minimal assumptions about the hardware was chosen in order not to limit its modeling capacity by mistaken beliefs about the hardware.
We use the expressive power of transformers, particularly their ability to learn from sequences, to explore the importance of ordering effects.
More specifically, a standard transformer encoder of 3 stacked layers with learned positional embeddings followed by a regression head to represent the relationship between dot product input vectors and scalar output was used. Each transformer encoder layer has a feed-forward dimension of 2048 and 8 attention heads. An illustration of the model setup can be seen in Figure~\ref{fig:transformer_schema}.

As discussed in Section \ref{subsec:bss2}, analog computations on BSS-2 include device-specific static variations as well as dynamic noise.
Furthermore, it could be shown that the aforementioned uncertainties are not only device-specific, but also neuron-specific.
In fact, each neuron will exhibit static variations and dynamic noise which differs significantly to the variations and noise of its neighboring neurons.
Therefore, we chose again to model each of the 256 accelerator neurons explicitly by its own transformer, resulting in a \emph{transformer set} covering the static variations of each neuron of an analog core.  

Additionally, to address the dynamic noise present on BSS-2, which is not possible with a purely deterministic model, we add noise sampled from a zero-mean Gaussian distribution, where the variance was measured for each neuron individually by performing a set of matrix multiplications multiple times.
This noise representation is identical in the \emph{white-box model} and a more detailed discussion is available in~\cite{klein2021item}.
By combining the neuron-specific \emph{transformer set} and the Gaussian noise, it is possible to emulate a matrix multiplication on the analog core, where each neuron performs a dot product between the input vectors and its weight matrix column.

\subsection{Dataset Composition}\label{subsec:dataset}
Building upon our previous work, a variety of measurements were performed in order to capture the characteristics of the analog hardware and to construct a dataset to train the neuron-specific \emph{transformer set} on.

\emph{Row-wise measurements} can reveal each neuron's output for all possible inputs at each position along the synapse array.
In this case, saturation and other effects are minimized, since only the specific row in the synapse array is active, and the electrical signal strength is low.
This measurement is designed to capture the specific non-linearities for each synapse independently. 

With a \emph{full-range measurement}, occurring saturation effects and inter-row interactions can be captured by varying all possible activations, weights and input vector sizes.  

Furthermore, a measurement with \emph{random values}, originally designed to measure the noise of each neuron, was also added to the dataset.
Here, the randomly-filled input and weight matrices can reveal device behavior for different input sizes in the smaller/middle part of the allowed output range, in contrast to the constant vectors and saturation cases found in the \emph{full-range measurement}.

To not only include constructed dot products, but also "real-world" data that better represents the actual usage of BSS-2, the inputs, weights and results of matrix multiplications from an ML task trained on device are used to enrich the dataset.
Therefore we used data from \emph{hardware-in-the-loop} training of the same model and keyword spotting task as described in Section~\ref{sec:results}. 
This model was first pretrained and then deployed on BSS-2, where \emph{hardware-in-the-loop} training took place for 300 epochs.
The data was then collected from the whole training process equidistantly to capture the states of a model in varying degrees accustomed to the device's peculiarities.

Overall, each sample in the combined dataset has two vector inputs and a scalar ouput representing the accelerator's calculated output of the dot product between the two input vectors. Consequently, the \emph{transformer set} is able to learn the behavior of BSS-2 when computing dot products.

\subsection{Experimental Setup to Evaluate the Importance of Ordering}

We trained two full \emph{transformer sets} on the described dataset of dot products, where the first set receives the input vectors in the correct order and will be referred to as \emph{ordered transformer set} in the following. The second set receives the input operands randomly permuted (but in unison) and will be referred to as \emph{non-ordered transformer set}. Each transformer in both sets was trained for 300 epochs with a learning rate given by $lr=lr\_base*0.999^{epoch}$, with a base learning rate of 1e-3. With these \emph{transformer sets}, we aim to show that the \emph{ordered transformer set} is able to capture an essential part of the chip’s behavior that the \emph{non-ordered transformer set} is lacking.

To validate this hypothesis, an ML model was trained with either one of the \emph{transformer sets} and then deployed on BSS-2 for inference.
This experiment is designed to reveal how much the quality of the \emph{transformer sets} as a proxy for the analog matrix multiplication depends on correct ordering information.
\section{Training Methods and Results}
\label{sec:results}

\subsubsection{Keyword Spotting Task}

\begin{wrapfigure}[16]{r}{0.22\linewidth}
		 \vspace{-4\baselineskip}
		 \centering
		 \includegraphics[width=0.22\textwidth]{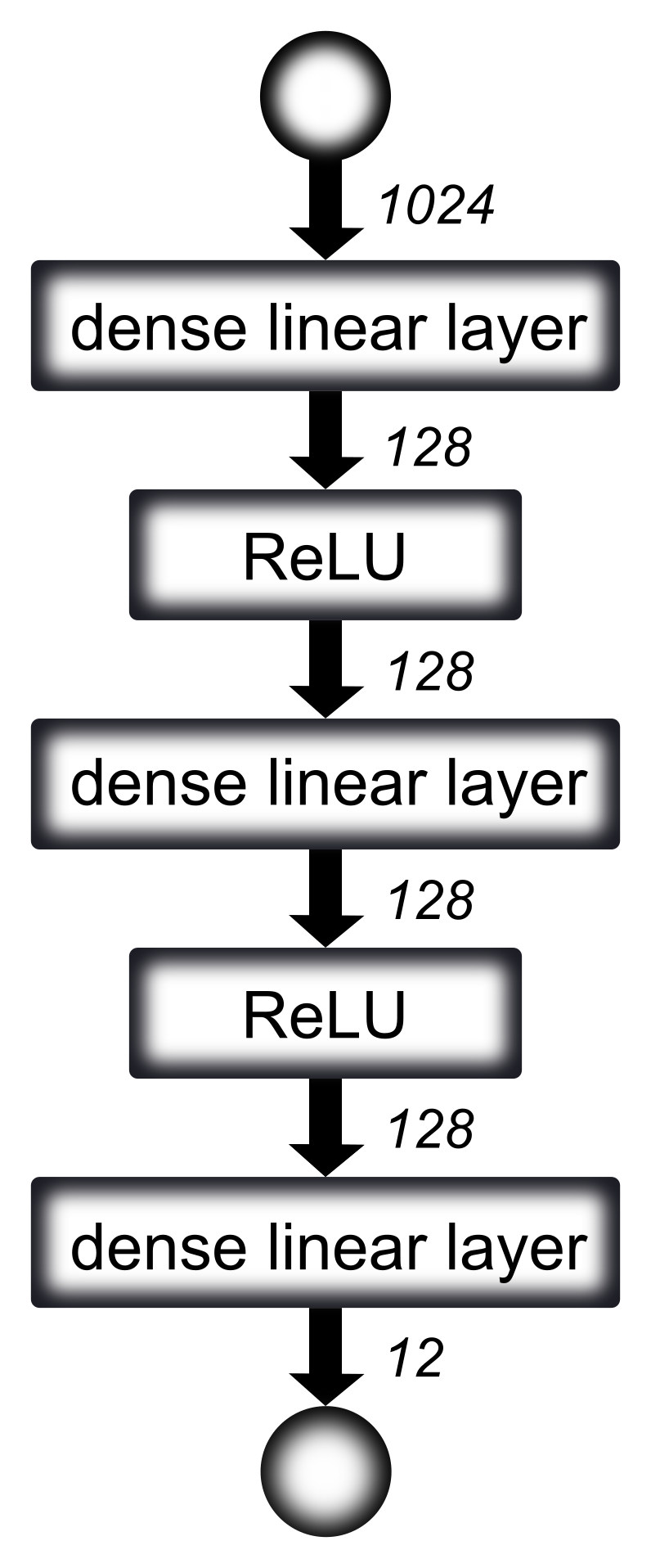}
		 \caption{Keywordspotting model}
		 \label{fig:keywordspottingmodel}
\end{wrapfigure}

As an exemplary task of sufficient difficulty to validate our hypothesis, keyword spotting on the GSC data set~\cite{warden2018speechcommands} was used.
In order to keep the keyword spotting model's complexity inline with the analog device's limitations, a multi-layer perceptron with log-mel filter \cite{mermelstein1976melspec}, as illustrated in Figure~\ref{fig:keywordspottingmodel}, was chosen over other architecture options like CNNs, LSTMs or transformers.

\subsubsection{Training Setup}
To evaluate the importance of ordering, the \emph{transformer sets} were used to emulate the hardware in a retraining phase.
The procedure is as follows: First, the keyword spotting model was normally trained in full precision for 300 epochs, without any considerations about the analog device yet.
Then, with the two \emph{transformer sets} as proxy for the analog device, the full-precision model was retrained with either of the two sets in the forward path for 300 more epochs.
Here, it could optimize for the necessary quantization, device-specific noise as well as the non-linearities---and in the case of the \emph{ordered transformer set}, ordering information---that these transformers were able to learn and represent.
Experiments with the \emph{white-box model} have shown that increasing the level of added Gaussian noise during training from 0\% to 100\% of the real noise level proved useful for the model's robustness and was therefore repeated in this setting. 
Finally, after the retraining with either of the \emph{transformer sets}, inference on the analog device itself was performed. 

\begin{table}[h!]\renewcommand{\arraystretch}{1.0}
	\centering
	\small
	\begin{tabular}{c|  >{\centering}p{15mm}|| >{\centering}p{15mm} |  >{\centering\arraybackslash}p{15mm}}
	\multirow{2}*{\bfseries Method} & \bfseries MSE  & \multicolumn{2}{c}{\bfseries GSC Test Acc.}	\\ \cline{3-4}
	 &  \bfseries to BSS-2 & \bfseries SW & \bfseries BSS-2 \\ \hline 

	Ordered Transformer Set & 0.5 & 76.6\% & 53.4\% \\
	Non-ordered Transformer Set & 0.9 & 77.8\% & 44.5\% \\
	White-box Model & -  & 76.9\% & 41.2\% \\
	\end{tabular}
	\caption{Comparison of the three different hardware models. Their ability to model the hardware in the created dataset correctly (MSE) and test accuracy for a GSC model trained with the respective hardware models and evaluated on the real analog device. The transformer set with ordering information performs clearly better, in representing the hardware directly and via retraining.}
	\label{tab:results}
	\vspace{-3\baselineskip}
\end{table}

\subsubsection{Results}
Table \ref{tab:results} compares the three different hardware modeling approaches.
The \emph{ordered transformer set} is able to represent the accelerator's behavior better, as can be seen directly by the lower MSE on a held-out test set of the dataset.
Additionally, the higher accuracy of the corresponding retrained GSC model with \emph{model-in-the-loop} and evaluated on the analog device further emphasizes this.
Compared to the \emph{white-box model}, we can observe a small improvement with the \emph{non-ordered transformer set}, which hints that the much larger model was able to learn some effects which were not represented within the explicit model.
However, this effect is small compared to the notable improvement of the \emph{ordered transformer set} over the other two approaches.
This proves our hypothesis that ordering matters and analog accelerators have to be considered non-associative.

\begin{table}[!h]\renewcommand{\arraystretch}{1.0}
	\centering
	\small
	\begin{tabular}{c|c|c|c}
	\multirow{2}*{\bfseries Method} &\bfseries HW Model &\bfseries BSS-2& \bfseries Accuracy on \\
	 & \bfseries Retraining Eps. & \bfseries Retraining Eps. & \bfseries BSS-2  \\ \hline 
	Without HW Model & - & 350 & 69.6\% \\
	Ordered Transformer Set & 300 & 50 &71.8\% \\
	Non-ordered Transformer Set & 300 & 50& 71.6\% \\ 
	White-box Model & 300 & 50&  71.9\% \\ 
	\end{tabular}
	\caption{GSC test accuracy of the models after hardware-in-the-loop retraining.}
	\label{tab:hwLoopResults}
	\vspace{-2\baselineskip}
\end{table}

It must be noted, that despite the important finding that ordering is of importance, all representations for BSS-2 still could not compete with retraining with the hardware itself, compare Tables~\ref{tab:results} and~\ref{tab:hwLoopResults}. 
Training for the same amount of epochs with \emph{hardware-in-the-loop} outperforms a \emph{ordered-transformer-set-in-the-loop} by $\sim 16\%$.
However, after 50 epochs of \emph{hardware-in-the-loop} training on top of \emph{model-in-the-loop} training with any of the three representations, all GSC models slightly outperform exclusive training on BSS-2 with the same amount of epochs.
This is inline with our previous observations~\cite{klein2021item}.
The 50 retraining epochs with the real hardware are also sufficient for the models without correct ordering information to incorporate the missing information.
Overall, the results show that considering the ordering effects is a big step towards modeling analog hardware correctly, but there are still other effects at play that have to be described and understood.

\section{Summary and Outlook}

In summary, we can report that the hypothesis on the non-associativity of analog computations based on CMOS holds true, actually to an extend that neglecting such ordering effects of input operands results in substantially reduced accuracy for machine-learning tasks.
We thus propose to always include ordering effects when it comes to modeling CMOS-based analog hardware for tasks related to machine learning.

For a keyword-spotting task, a direct comparison of hardware models with and without ordering effects shows that prediction accuracy improves by $9-12\%$, depending on the baseline.
This is a substantial step ahead and gives evidence on the presence and importance of non-associative effects.
Furthermore, the ablation study showed that when we remove ordering effects from the hardware model, the resulting accuracy reverts to previous results based on a \emph{white-box model}, which also did not consider ordering effects.

However, as the achieved accuracy is still far from the accuracy achieved with training on real hardware ($72\%$), there is a strong need for future work.
Since an \emph{ordered transformer set} as presented in this work shall be able to capture all effects which are static in nature, such as non-linearity, saturation and ordering, we hypothesize that there are dynamic effects left which are yet to be uncovered.
Future work will focus on such dynamic effects, in particular related to time-dependent nonlinearities and noise, as well as extend to more learning tasks including diverse model architectures and data sets.

\section*{Acknowledgements}

We greatly acknowledge the extensive and continuing support of the Electronic Visions Group, led by Johannes Schemmel, from Heidelberg University for valuable discussions and assistance in all matters, and for providing access to BSS-2.

This work is part of the COMET program within the K2 Center “Integrated Computational Material, Process and Product Engineering (IC-MPPE)” (Project No 886385), and supported by the Austrian Federal Ministries for Climate Action, Environment, Energy, Mobility, Innovation and Technology (BMK) and for Labour and Economy (BMAW), represented by the Austrian Research Promotion Agency (FFG), and the federal states of Styria, Upper Austria and Tyrol.

\bibliographystyle{splncs04}
\bibliography{references}

\end{document}